\documentclass[aps,pra,showpacs,twocolumn]{revtex4-1}

\usepackage{hyperref}
\usepackage{graphicx}
\usepackage{amsmath}
\usepackage{amsfonts}
\usepackage{amssymb}
\usepackage{times}
\usepackage{subfigure}
\usepackage[usenames,dvipsnames]{color}
\usepackage{bm}
\everymath{\displaystyle}
\hypersetup{
  colorlinks=true,
  citecolor=blue,
  linkcolor=blue,
  urlcolor=blue}
\begin{document}
\title{Finite temperature expansion dynamics of Bose-Einstein condensates}

\author{Arko Roy}
\author{D. Angom}
\affiliation{Physical Research Laboratory,
             Ahmedabad - 380009, Gujarat,
             India}

\date{\today}


\begin{abstract}

We explore the effects of finite temperature on the dynamics of Bose-Einstein 
condensates (BECs) after it is released from the confining potential. In 
addition, we examine the variation in the expansion dynamics of the BECs as 
the confining potential is transformed from a multiply to a simply connected 
geometry. To include the effects of finite temperatures we use the frozen 
thermal cloud approximation, and observe unique features of the 
condensate density distribution when released from the confining potential. We
find that at $T\neq 0$, during the initial stages of expansion, the multiply 
connected condensate has more pronounced interference rings compared to the 
case of zero temperature. Such difference in the dynamical evolution is 
also evident for simply connected condensates.

\end{abstract}





\maketitle

\section{Introduction}

The time-of-flight measurement is an important experimental technique to detect 
Bose-Einstein condensation in dilute atomic gases, and probe their other 
properties as well. In this technique the atoms are left to expand by
switching off the external confining potential, and the atoms are then 
imaged with optical methods. The technique has been used to observe a 
plethora of diverse phenomena in the Bose-Einstein condensates (BECs) of dilute
atomic gases. Few examples are the experimental observation of the 
non-equilibrium many-body phenomenon based on matter wave interference
patterns~\cite{smith_13}, and the observation of thermally activated vortex 
pairs in a quasi-2D Bose gas leading to a crossover from a Berezinskii-
Kosterlitz-Thouless phase to a vortex-free BEC~\cite{choi_13}. In fermionic
atomic species, the technique has been used to probe the
superfluidity of strongly interacting Fermi mixtures~\cite{hara_02}
and, to measure the $p$-wave Feshbach resonances for fermionic
atoms~\cite{regal_03}.

  Toroidal condensates, which are multiply connected BECs, are near ideal
systems to study phenomena related to persistent superflows. In these systems
too, the time-of-flight measurements play an important role to detect and
probe the  phenomena of interest. In experiments, toroidal condensates have been
obtained with the use of harmonic potential in combination with a Gaussian
potential~\cite{ryu_07},
Laguerre-Gaussian beams~\cite{ramanathan_11,moulder_12,curtis_03,beattie_13},
combination of an RF-dressed magnetic trap with an optical
potential~\cite{heathcote_08,morizot_06},
magnetic ring traps~\cite{sauer_01,gupta_05,arnold_06,sherlock_11},
time-averaged ring potentials~\cite{henderson_09,bell_16}, coincident red
and blue detuned laser beams~\cite{marti_15}, and employing digital micromirror
devices ~\cite{kumar_16}. The toroidal or ring condensates also serve as 
model-systems in the field of atomtronics. These systems have been used to 
implement superconducting quantum interference devices 
(SQUIDs)~\cite{ryu_13,mathew_15}, and  phase-slips in rf SQUID have been 
modelled 
by employing blue-detuned laser beam along the axis of toroidal 
condensates~\cite{wright_13}. In both of these experiments, the
phase-difference along the condensate has been measured from the free
expansion dynamics. It is found that for toroidal BECs with finite circulation
the images after expansion have a central hole whose area is proportional to 
the winding number. On the other hand, there is finite density at the center
when there is no circulation~\cite{wright_13,murray_13}. These are well 
understood and in good agreement with theoretical simulations using zero
temperature time-dependent Gross-Pitaevskii (GP) equation~\cite{murray_13}.

For the present work, we consider a toroidal BEC obtained with a confining
potential consisting of a harmonic and Gaussian potential~\cite{ryu_07}.
This configuration offers the possibility to probe the effects associated 
with the transition from multiply to simply connected BEC due to relative 
shift in the component trapping potentials. In experiments, this is an
important consideration since the trap centers, extremum of the harmonic
and Gaussian trapping potentials, never coincide. This is due to  gravitational 
sagging, and deviations of the optical elements and external fields
from perfect alignment. In addition, it is worth mentioning that,
in experiments, the drift due to thermal cycling alone transforms the
multiply to simply connected BEC~\cite{ryu_07}. These experimental
realities establish the need to theoretically probe the effects of
geometry on the expansion dynamics of a quasi-2D BEC. To validate
the experimental results, and considering the deviations from an ideal
case, it is also pertinent to examine the role of thermal fluctuations on 
the expansion dynamics. So that, it is possible to distinguish and identify
the effects of relative shift in the trapping potential, and those
emerging from the thermal fluctuations.

 To examine the thermal fluctuations in the toroidal BEC as a function of the 
separation between the trap centers we use the Hartree-Fock-Bogoliubov 
theory with Popov (HFB-Popov) approximation. The increase in the separation of
the trap centers induces a topological transformation in the condensate
density profiles. The condensate density profile is modified from toroidal or 
multiply connected to a bow-shaped or simply connected geometry, the profile
of the quantum fluctuations also exhibit a similar transformation. The thermal 
fluctuations, in contrast, remain multiply connected~\cite{roy_16}. These 
differences in the structure of the condensate and thermal density profiles
affect the dynamics of the condensate cloud during expansion at finite
temperature. We examine this in detail using frozen thermal cloud approximation. 
It is to be mentioned here that, previous works using the classical field 
approximation have shown that thermal fluctuations affect the radial and axial 
condensate widths during expansion~\cite{zawada_08,gawryluk_10}.


\section{Theoretical methods}
The grand-canonical Hamiltonian of an interacting quasi-2D BEC system
is
\begin{eqnarray}
  \hat{H} &=& \iint dx dy\,\hat{\Psi}^\dagger(x,y,t)
        \bigg[-\frac{\hbar^{2}}{2m}\left(\frac{\partial ^2}{\partial x^2} +
        \frac{\partial ^2}{\partial y^2}\right)
        + V(x,y)\nonumber\\
      &&  -\mu + \frac{U}{2}\hat{\Psi}^\dagger(x,y,t)\hat{\Psi}
        (x,y,t)\bigg]\hat{\Psi}(x,y,t).
\label{hamiltonian} 
\end{eqnarray}
where $\hat{\Psi}$ and $\mu$ are the Bose field operator of a scalar BEC, 
and the chemical potential, respectively. We consider an external confining 
potential of the form 
$V(x,y) = (1/2)m\omega_x^2(x^2 + y^2 + \lambda^2z^2) + U_0e^{{-[(x-\Delta_x)^2 
+ y^2}]/2\sigma^2}$, which is a 
superposition of the harmonic oscillator and Gaussian potential with
strength $U_0$. This choice of the potential parameters implies that the
aspect ratio in the transverse direction $\alpha = \omega_y/\omega_x=1$.
The case of $U_0=0$, $\Delta_x=0$ then produces a symmetric 2D harmonic 
trap, whereas $U_0\gg0$ makes it a toroid. Let $\Delta_x$ represent the 
separation 
between the centers of the harmonic and Gaussian potentials, we
can consider the separation as along the $x$-axis through an appropriate 
rotation of the coordinates. This separation accounts for the
non-coincidence of the trapping potential centers. This
deviation is natural, since in experiments trap centers never coincide 
because of gravitational sagging, and optical axes are not perfectly aligned. 
In a quasi-2D system, as $\omega_z\gg\omega_x$ the condensate is 
considered to be in the ground state along $z$, and the excitations along 
the transverse direction only contribute to the dynamics. The atoms of the 
bosonic species with mass $m$ and scattering length $a$ interact 
repulsively through the $s$-wave binary collisions with strength 
$U = 2 a\sqrt{2\pi\lambda}$.


\subsection{Gapless Hartree-Fock-Bogoliubov-Popov formalism}

To compute the equilibrium density profiles of BEC at finite temperatures,
we use the gapless HFB-Popov theory. In this theory the Bose field operator
$\hat\Psi$ is decomposed into a condensate part represented by
$\phi(x, y, t)$, and the fluctuation part denoted by $\tilde\psi(x,y,t)$.
That is $\hat{\Psi} = \phi + \tilde\psi$, and the condensate part 
$\phi(x, y, t)$ solves the generalized GP equation 
\begin{equation}
  \hat{h}\phi + U\left[n_{c}+2\tilde{n}\right]\phi = 0.
  \label{gpe1s}
\end{equation}
In the above equation the single-particle or the non-interacting part of the 
Hamiltonian is $\hat{h}= (-\hbar^{2}/2m)\left(\partial ^2/\partial x^2
                 +\partial ^2/\partial y^2\right) + V(x,y)-\mu$ 
with $n_{c}(x,y)\equiv|\phi(x,y)|^2$,
$\tilde{n}(x,y)\equiv\langle\tilde{\psi}^{\dagger}(x,y,t)
\tilde{\psi}(x,y,t)\rangle$, and $n(x,y) = n_{c}(x,y)+ \tilde{n}(x,y)$
as the local condensate, non-condensate or thermal, and total density,
respectively. To determine the non-condensate density, the fluctuation
operator $\tilde{\psi}$ is represented through a superposition of 
Bogoliubov quasiparticle as
\begin{eqnarray}
   \tilde{\psi} &=&\sum_{j}\left[u_{j}(x,y)
    \hat{\alpha}_j(x,y) e^{-iE_{j}t/\hbar}
   - v_{j}^{*}(x,y)\hat{\alpha}_j^\dagger(x,y) e^{iE_{j}t/\hbar}
    \right], \nonumber\\ 
\label{ansatz}
\end{eqnarray}
with $j$ denoting the energy eigenvalue index of a quasiparticle mode
having energy $E_j$. The quasiparticle annihilation (creation) 
operators $\hat{\alpha}_j$ ($\hat{\alpha}_j^\dagger$) satisfy the usual Bose 
commutation relations. The functions $u_j$ and $v_j$ are the Bogoliubov 
quasiparticle amplitudes corresponding to the $j$th energy eigenstate, and 
solves the following pair of coupled Bogoliubov-de Gennes (BdG) equations
\begin{subequations}
\begin{eqnarray}
(\hat{h}+2Un)u_{j}-U\phi^{2}v_{j}&=&E_{j}u_{j},\\
-(\hat{h}+2Un)v_{j}+U\phi^{*2}u_{j}&=& E_{j}v_{j}.
\end{eqnarray}
\label{bdg1}
\end{subequations}
\noindent Following these definitions, the thermal or the non-condensate 
density at temperature $T$ is 
\begin{equation}
 \tilde{n}=\sum_{j}\{[|u_{j}|^2+|v_{j}|^2]N_{0}(E_j)+|v_{j}|^2\},
  \label{n_tilde}
\end{equation}
where $\langle\hat{\alpha}_{j}^\dagger\hat{\alpha}_{j}\rangle = (e^{\beta
E_{j}}-1)^{-1}\equiv N_{0}(E_j)$ with $\beta=1/k_{\rm B} T$, is the Bose
factor of the $j$th quasiparticle state with energy $E_j$ at temperature $T$.
More details of the derivations, and numerical scheme to solve the 
generalized GP and coupled Bogoliubov-de Gennes (BdG) equations 
self-consistently are given in our previous works
~\cite{roy_16,roy_15a,roy_14,roy_14a}.


\subsection{Frozen thermal cloud approximation}

To study the dynamics of BEC at finite temperatures, we solve the
time dependent GP equation with frozen thermal cloud approximation. In this 
approximation the dynamics of the thermal cloud is ignored, and the 
condensate atoms move in the presence of a static cloud of non-condensate 
atoms which are in a state of thermal equilibrium with the initial state, and 
obeys Bose-Einstein distribution function. This, in the expansion dynamics
of the condensate after the removal of the trapping potential, is equivalent 
to introducing a perturbation potential to the condensate atoms when $T > 0$. 
The total number of atoms during the evolution is conserved, in other words
there is no transfer of atoms between the condensate and thermal clouds. Thus 
the dynamics of the order parameter $\phi$ of the BEC under frozen thermal 
cloud approximation follows the equation
\begin{eqnarray}
 i\hbar\frac{\partial \phi}{\partial t} =
 \left[-\frac{\hbar^2}{2m}\nabla_{\perp}^2 + 2U\tilde{n} +
 Un_c\right]\phi(x,y,t).
\label{fta} 
\end{eqnarray}
Here $\nabla_{\perp}^2 = \partial^2/\partial x^2 + \partial^2/\partial
y^2$, and $U$ is interaction strength strength as defined earlier. The time 
independent interaction with the thermal density $\tilde{n}$
acts as an effective potential in which the condensate moves. To probe the
region of validity of this approximation we compute the velocity of the
thermal cloud through the computation of root mean square value of
wave vector
\begin{eqnarray}
 k^{\rm rms}_{\rm th} = 
\left\{\frac{\iint dk_x dk_y (k_x^2 + k_y^2){\tilde n}(k_x,k_y)}
{\iint dk_x dk_y {\tilde n}(k_x,k_y)}\right\}^{1/2}.
\end{eqnarray}
Thus the root mean square velocity of the thermal atoms is given by
$v^{\rm rms}_{\rm th} = \hbar k^{\rm rms}_{\rm th}/m$. On switching off the 
trap, that is, during free expansion of the condensate cloud in time interval 
$\delta t_c$, the change in the radial mean size of the thermal cloud 
$\delta r^{\rm rms}_{\rm th} = v^{\rm rms}_{\rm th}\times \delta t_c$. If
during this time $\delta t_c$, if $\delta r_c$ is the change in the radial 
mean size of the condensate, then the frozen thermal cloud approximation
is valid when $\delta r^{\rm rms}_{\rm th}\ll \delta r_c$.


\subsection{Free expansion of BEC}

To simulate the time-of-flight density evolutions, we obtain the finite 
temperature equilibrium solutions of a single species BEC by solving
Eqns.~\ref{gpe1s} and \ref{bdg1} self-consistently. Using the resulting
condensate density implies that it has less number of atoms than the 
condensate density at $T=0$, we then evolve Eq.~\ref{fta} in real time using
the split-step Crank-Nicolson algorithm on a 2D spatial grid. As mentioned 
earlier, the dynamics of $\tilde{n}$ is neglected under this approximation, 
and we are interested at looking at the non-equilibrium behaviour of the 
condensate cloud in the presence of a static thermal background. The 
dimensions of the surface area considered is  chosen large enough to 
eliminate effects of bounces from the boundaries during the free-expansion.
Immediately, after the release the BEC expands along the radial direction
as the inter-atomic interaction potential energy is converted to kinetic 
energy. For $U_0\gg0$ and $\Delta_x=0$, the stationary cloud is azimuthally 
symmetric and multiply connected. However, when $\Delta_x\neq0$, the azimuthal 
symmetry is broken, and the density evolutions are dramatically different.


\section{Results and Discussions}

\begin{figure}[t]
 \includegraphics[width=8.0cm]{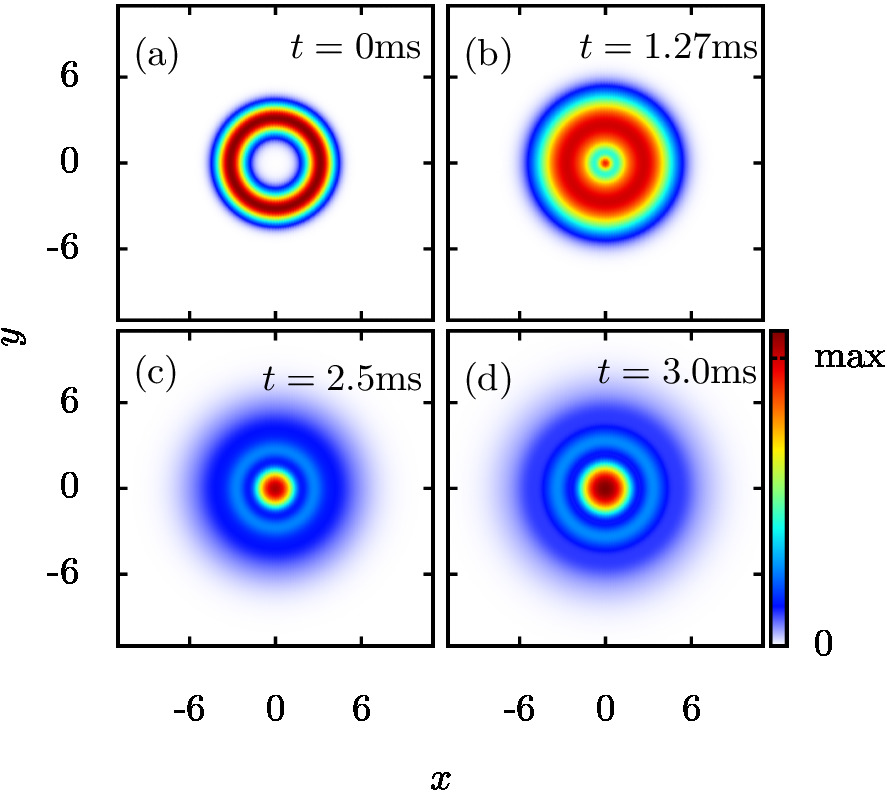}
  \caption {(Color online) Plots showing condensate density profiles at 
            different instants of time after release from the trap 
            with $\Delta_x=0$ and $T=0$. Density is measured in units 
            of $a_{\rm osc}^{-2}$.
            }
  \label{exp0}
\end{figure}

\subsection{Zero temperature Expansion Dynamics}

The typical dynamical evolution of $ n_{c}(x,y)$ after releasing from a 
toroidal trapping potential is shown in Figs.~\ref{exp0}. As to be expected
at $t=0$, $ n_{c}(x,y)$ is toroidal in shape. After switching off the 
trapping potential, the release of the repulsive interaction energy creates
two distinct components in the evolution. The first consists of the inner 
portion of the condensate which implode towards the origin, and the 
second component is the the outer portion which moves outward. For the 
imploding component, atoms along the toroid fill up the hole initially 
created due to the Gaussian potential. The atoms accumulate at the center
with increasing repulsive interaction energy till $n_c(0,0)$ reaches a 
critical density. After attaining the critical density, the atoms expand 
outwards or explodes. These are the general features of the evolution after
the removal of the trapping potential. However, different values of
$\Delta_x$ lead to significant differences in the dynamics. In other words,  
separation
between the minima of the oscillator potential, and the maxima of the 
Gaussian obstacle potential modify the expansion dynamics. In experiments 
$\Delta_x\neq 0$ due to gravitational sagging or due to imperfect alignment of 
the optical axes.
\begin{figure}[t]
 \includegraphics[width=9.0cm]{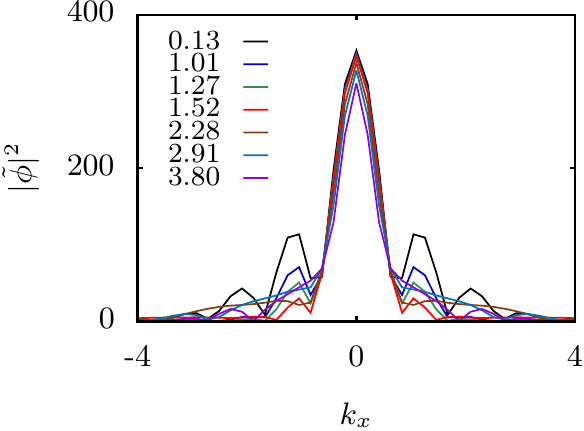}
 \caption { (Color online) Plots showing a cut through of the momentum 
           distribution of the atoms at different instants of time after 
           the release of the trap with $\Delta_x=0$ and $T=0$.
          }
\label{ftx1d}
\end{figure}


\subsubsection{Coincident potential centers}
For the coincident trap centers, there is a prominent self-interference 
between the expanding outer portion of the condensate, and the condensate 
cloud resulting from the repulsive explosion at the center. The interference 
occurs due to higher repulsive energy, and hence the velocity of expansion, 
accumulated at the center during the implosion. For example, in the case 
of $U_0=15\hbar\omega_x$ the central or maximum density 
reaches $0.05 a_{\rm osc}$. This is more than double the maximum density, which 
is $\approx 0.02 a_{\rm osc}$ along the toroidal axis, in presence of the
trapping potential. The self-interference results in the appearance of rings 
in the condensate density profile, and the there is a dynamical change in the 
pattern as the condensate expands. 

To probe the velocity distribution and release of the repulsive energy
during the expansion, we compute the Fourier transform of the condensate 
density in the $k_x-k_y$ space. This in essence reveals the kinematic
aspects of the condensate density profiles, and relates to the dynamics
prior to the experimental observations based on time-of-flight(TOF) imaging. 
With time, the momentum distribution of the condensate atoms shrinks or 
broadens depending upon the release of the potential energy from the 
inter-atomic interaction or the mean field energy. The growth in the
condensate density at the center of the trap is accompanied by a suppression 
of atoms with zero momentum. During the course of free expansion, the low-lying 
momentum peaks are suppressed, and the momentum profiles become broader. That is
to say, the release of the repulsion energy leads to atoms having higher 
momenta. This is also evident from the plot of $k_x$ in Fig.~\ref{ftx1d}. 
However, as the the central density reaches a critical value, which is 
sufficient enough to induce an expansion, the low-lying momentum peaks get 
washed out. Hence, there equal number of atoms for a wide range of momentum, 
and the interference rings start to appear in the condensate. At later times, 
the interference pattern becomes more prominent, and the atoms in the condensate
show well defined distribution peaks in momentum. This is apparent from the 
presence of the peaks with higher harmonics in the Fourier spectrum.

\begin{figure}[t]
 \includegraphics[width=8.0cm]{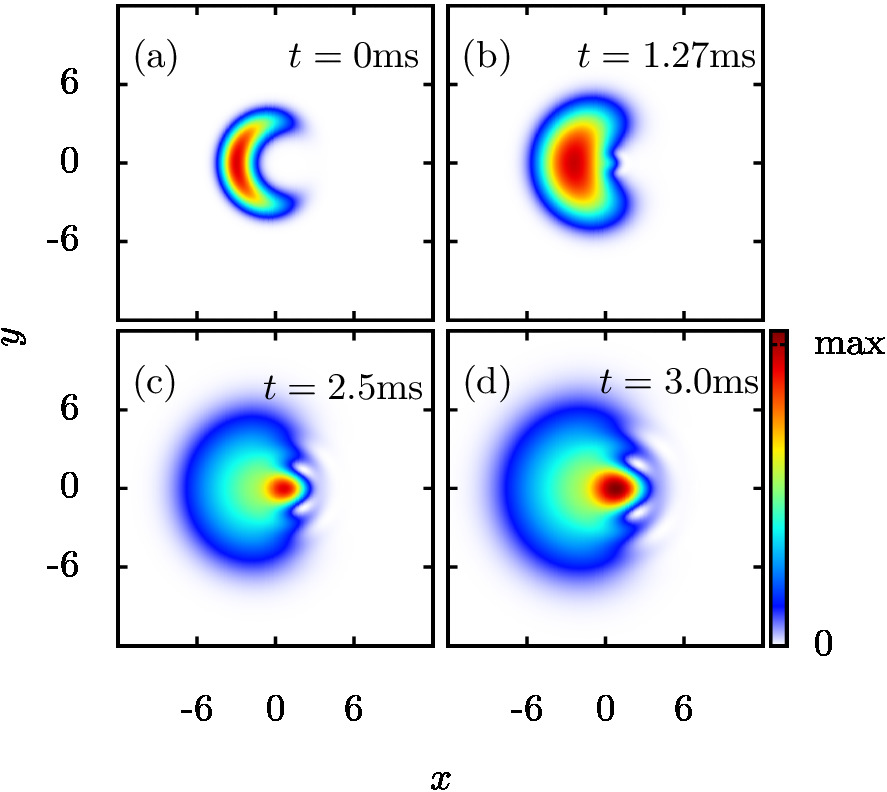}
 \caption {(Color online) Plots showing condensate density profiles at
            different instants of time after release from the trap
            with $\Delta_x=0.4 a_{\rm osc}$ and $T=0$. Density is measured 
            in units of $a_{\rm osc}^{-2}$.
          }
\label{exp04}
\end{figure}

\begin{figure}[t]
 \includegraphics[width=8.0cm]{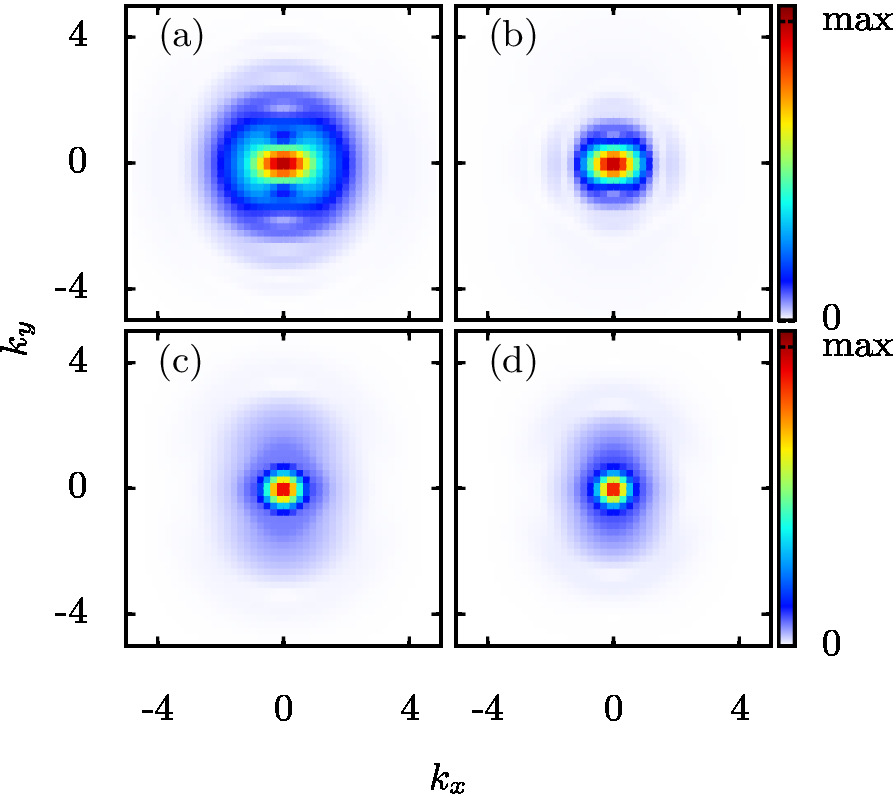}
 \caption {Plots showing the momentum distribution of the condensate density 
           profiles at different instants of time after release from the trap
           with $\Delta_x=0.4 a_{\rm osc}$ and $T=0$.
          }
\label{ftx04}
\end{figure}


\subsubsection{Non-coincident potential centers}

When $\Delta_x\neq 0$, the rotational symmetry of the potential is
broken, and the condensate cloud undergoes a transformation in topology.
As $\Delta_x$ is increased, the condensate density profile assumes a bow-shaped
geometry and becomes simply-connected. However, it must be mentioned here 
that, the simply connected profile is different from the pancake shaped 
condensate when $U_0=0$ and $\Delta_x=0$. Hence, a difference in the dynamics 
of the free expansion is to be expected. At $\Delta_x = 0.4$, the equilibrium 
density profile of the condensate is shown in Fig.~\ref{exp04}(a). From the 
figure it is evident that the density gradient along the periphery is 
non-uniform. As the trapping potential is removed, the condensate expands 
with anisotropic density and velocity distribution, and the atoms fill up 
the central void. As the condensate expands, the outer component continues to 
bear the signature of the initial density anisotropy. The inner component,
on the other hand, on expansion is close to rotational symmetry.
\begin{figure*}[t]
 \includegraphics[height=8.5cm]{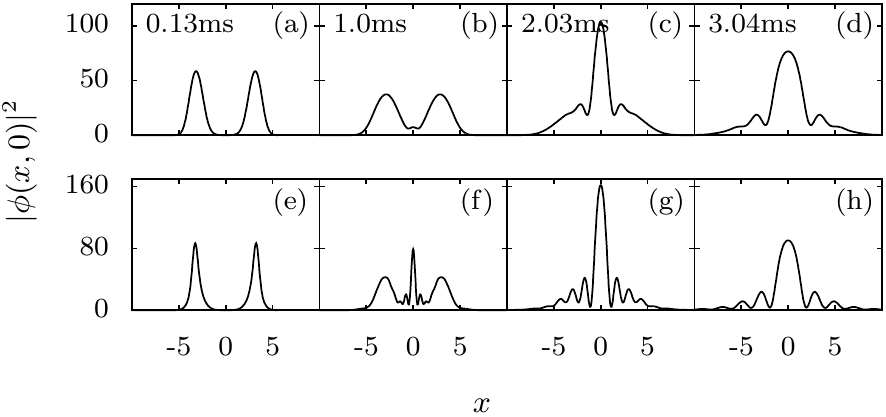}
 \caption{(Color online) Cut through of condensate density
           profiles in different instants of time after release from the
           trap with $\Delta_x=0$ at (a)-(d) $T=0$, and (e)-(f) $T=10$nK.
           Density is measured in units of $a_{\rm osc}^{-2}$.
         }
\label{den1d}
\end{figure*}

As mentioned earlier, to probe the momentum distribution of the released 
atoms, we perform the Fourier transform of the condensate
density in the $k_x-k_y$ space. As density distribution is anisotropic along
the azimuthal direction, so is the momentum distribution. At the outset, when 
the density distribution assumes a banana-shaped structure with the density 
distribution constricted along the $x$-axis, and more extended along the 
$y$-axis, the momentum distribution of the atoms is more extended along the 
$k_x$. However, with the passage of time as the confining potential is 
switched there is a reversal of the extent of the momentum distribution. That 
is, at later times the momentum distribution along $k_y$ is larger than along
$k_x$. This is equivalent to stating that the condensate expands more along
the $x$-direction than $y$-direction in the position space. This is due to 
the release of the repulsion energy along the $x$-axis which, in presence of
the trapping potential, is associated with higher density.


\subsection{Finite temperature Expansion Dynamics}

We study the free expansion dynamics of the toroidal condensate at
$T=10$nK using the frozen thermal cloud approximation. For the system of
our interest, the change in the radial mean size of the condensate considered
during the evolution of 5 ms is $\approx 4.5 {\rm a_{osc}}$. The corresponding 
change in the thermal cloud size is $\approx 1 {\rm a_{osc}}$. Thus, the 
the condensate cloud may be assumed to move in a frozen thermal background. 
Based on our studies, we show that the presence of the thermal cloud brings 
about significant differences in the expansion dynamics of the condensate when 
released from the confining potential.


\subsubsection{Coincident trap centers}
At the outset when $t\approx 0.13$ms, that is soon after the confining
potential is switched off, we find a significant difference in the 
structure of the condensate density profiles at $T=0$ and $T=10$nK. This 
variation in the structure of the density profiles at the early time is 
reflected in the subsequent stages of evolution. It is evident from 
Fig.~\ref{den1d}(a) that the density profile is broader at $T=0$ than the
profile at $T=10$nK at $t\approx 0.13$ms. The sharp change in the
density gradient associated with a narrower density profile, as seen in 
Fig.~\ref{den1d}(e) at $T=10$nK, leads to a higher kinetic energy of the 
atoms. As a result, on switching off the confining potential the
atoms start to move inwards, and fill the void at the center with a faster rate 
till it reaches a density high enough so that the repulsive mean field energy
overcomes the kinetic energy. Along with this, due to the
interference between the incoming and outgoing atoms at a very early stage, 
rings begin to appear around the center of the trap at $T=10$nK 
after $t\approx1$ms of evolution. This is unique to finite temperature free 
expansion of toroidal condensate. With further evolution in time, such that the 
frozen thermal cloud approximation holds good, the density at the center 
continues to grow with the emergence of prominent ring structures. This is not 
the case for  $T=0$ where there are very few ring structures. During the final 
stages of the evolution, as the density reaches a critical value, due to the 
high repulsive energy between the atoms, it explodes and starts moving away 
from the center.


\subsubsection{Non coincident trap centers}
\begin{figure}[h]
 \includegraphics[width=8.0cm]{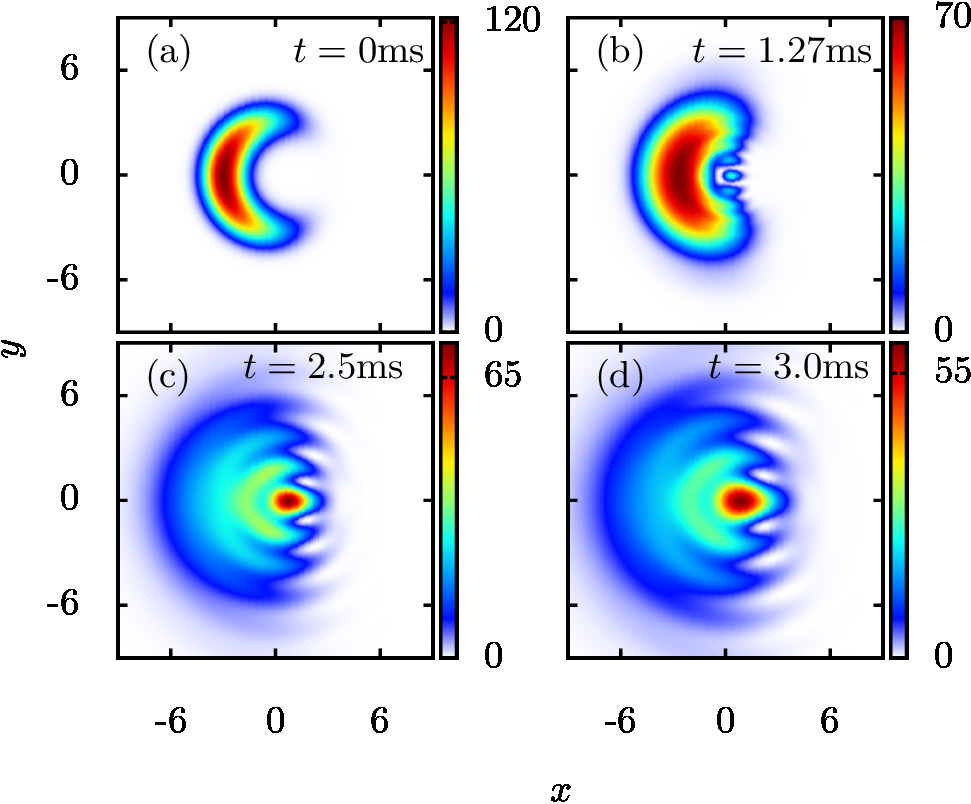}
 \caption{(Color online) Plots showing condensate density profiles at
           $T=10$nK in different instants of time after release from the trap
           with $\Delta_x=0.4 {\rm a_{\rm osc}}$. Density is measured in units
           of $a_{\rm osc}^{-2}$.
         }
 \label{exp04ft}        
\end{figure}

At finite temperatures, for the non-coincident trap centers, the expansion 
dynamics of the condensate after switching off the confining potential is 
very different from the zero temperature case. As the rotational symmetry of 
the trap is broken, the condensate density is anisotropic in the azimuthal 
direction. This makes the nature of the interaction between condensate, and 
thermal atoms different from the case with coincident trap centers. At zero 
temperature, we find that the velocity of the atoms around the center of the 
trap is higher than the atoms located along $(x=0,y=\pm \delta y)$. On 
switching off the confining
potential, the maxima of the condensate density expands much faster along the 
$+$ve $x$ direction, which at later times interfere with the atoms coming 
from $y=\pm \delta y$. However, at $T\neq 0$ during the free expansion, the 
additional repulsive interaction energy contribution from thermal atoms 
introduce different expansion features to the condensate density profiles. 
The velocity 
of the atoms along $y\in [y-\delta y, y + \delta y]$ are similar, and they form 
prominent ring like structures during the expansion. This is unique to 
finite temperature expansion dynamics and is evident from 
Figs. ~\ref{exp04}, ~\ref{exp04ft}.


\section{Conclusions}
\label{conclusions}

The present studies reveal distinct features of interference patterns in the 
expansion dynamics of the condensates in toroidal trap configuration at zero 
and finite temperatures. The offset between the centers of the harmonic and 
Gaussian confining potentials induces topological transformation to the 
condensate distribution. That is, it gets transformed from a multiply (toroidal)
connected to a simply connected geometry. At $T=0$, the simply connected 
condensate has an expansion that is geometrically very different from the 
multiply connected one. Unlike in the multiply connected case, during 
expansion, the simply connected condensate is devoid of interference rings. 
The central region, which initially collapse and later expands, has an 
ellipsoidal density distribution. However, the outer expanding part is 
semi-circular in structure, and with the center shifted from the center of 
both the harmonic and Gaussian trapping potentials. At $T \neq 0$,
because of the interaction between the condensate and thermal atoms, ring 
like self-interference structures are more prominent for both multiply, and 
simply connected condensates. The rings appear at the early stages of 
evolution.

\begin{acknowledgments}
We thank Mark Edwards, S. Pal, K. Suthar, S. Bandyopadhyay and R. Bai for 
useful discussions. The results presented in the paper are based on the
computations using Vikram-100, the 100TFLOP HPC Cluster at Physical Research
Laboratory, Ahmedabad, India.
\end{acknowledgments}

\bibliography{bec}{}
\bibliographystyle{apsrev4-1}

\end{document}